# The presence of interstellar turbulence could explain the velocity flattening in galaxies

Georgios H. Vatistas
Department of Mechanical and Industrial Engineering

Concordia University
1455 DeMaissoneuve Blvd. West
Montreal Canada
H3G 1M8

Expanding our previous work on turbulent whirls [1] we have uncovered a similarity within the similarity shared by intense vortices. Using the new information we compress the tangential velocity profiles of a diverse set of vortices into one and thus identify those that belong to the same genus. Examining the Laser Doppler Anemometer (LDA) results of mechanically produced vortices and radar data of several tropical cyclones, we find that the uplift and flattening effect of tangential velocity is a consequence of turbulence. Reasoning by analogy we conclude that turbulence in the interstellar medium could indeed introduce a flattening effect in the galactic rotation curves.

The physics of laminar vortices is relatively well known. Simple exact solutions of the Navier-Stokes derived by Rankine, Lamb-Oseen, Taylor, Burgers, Sullivan, and others exist. In the case of turbulent vortices our grasp on the subject reduces exponentially. Turbulent flows are highly complex phenomena and hence difficult to describe adequately. Even the subject of turbulent hydrodynamic vortices is a relatively unexplored frontier of classical hydrodynamics.

Not long ago, Ramasamy and Leishman [2] studied turbulent helicopter blade tip-vortices using improved instrumentation and experimental techniques. Their visualizations revealed that these whirls display the traditional path to turbulence. In region 1 ($0 < r < r_l$), see Fig. 1, the vortex is under laminar-like conditions. Approaching the neighborhood of maximum velocity ($V_{max}$), from the center of rotation, the flow enters a transition territory (region 2; $r_l < r < r_t$) when a critical local Reynolds number is reached at $r_l$. This state of affairs persists until a second critical local Reynolds number is attained at $r_t$ where the flowfield changes into the turbulent state, and remains throughout region 3. Most important, the accompanied high-resolution velocity data exposed the following particular behavior of the tangential velocity component. In region (3) the velocity decreases at a rate noticeably lower than of a laminar vortex. In retrospect however, this is not the first time that such flow behavior is encountered in classical fluid mechanics. It is a well know fact that in pipe and boundary layer flows (among others), the fluid velocity profile does flatten once turbulence sets-in.

It is known that for radii sufficiently larger than the core (radius of maximum velocity), the galactic tangential velocity lifts up, veering away from the laminar profile. The last manifestation is in disagreement with classical dynamics and is often rectified assuming the existence of nonbaryonic dark matter.

Interstellar gas clouds are turbulent and ionized [3]-[7], a subject that is poorly understood; presently under development [8]. The last attribute makes the treatment of any subject involving interstellar turbulence highly cumbersome. In this paper we take the route of analogy amongst turbulent vortices to explain the incongruity.

One of the classical methods of scientific inquiry is to examine phenomena through similarity. Two events are considered to be analogous, if their respective dimensionless equations describing these occurrences are of the same form. The reverse is also true. One of the most well known examples in fluid mechanics is the similar nature between shallow water hydraulics and the compressible gas flow [9]. The published visualization images of turbulent tip-vortices [2], [10] and images of cloud structure of tropical cyclones are in many respects similar to the basic coherent structure in galaxies. Therefore, there must exist an analogical relationship between these events.

In view of the fact that many physical details concerning turbulent flows are not available, a simpler approach is taken where only the bulk effects of turbulence are considered ignoring by necessity all the finer details. From a distance a galaxy appears to be a turbulent whirling motion taking place in a cosmic alphabet soup. The previously mentioned close resemblance between the tangential velocity profiles of mechanically produced turbulent vortices, the radar data of several geophysical vortices, and the galactic curves of several spiral galaxies led us into a similarity relationship common to all three categories. Then the newly uncovered similitude enabled us to show their kinship by compressing all profiles into one and then to deduce by analogy that high levels of turbulence in the medium could be the cause of the peculiar curve flattening in both tropical cyclones and galaxies. It is worth noting that Krishan [11], taking a theoretical line of reasoning, argued that a flow experiencing fluctuations in helicity, could give rise to coherent structures, which in addition to the familiar Kolmogorov spectrum leads into an energy plateau. He then concluded that it is unnecessary to invoke the presence of nonbaryonic dark matter in order to justify the flattening of the galactic rotational curves.

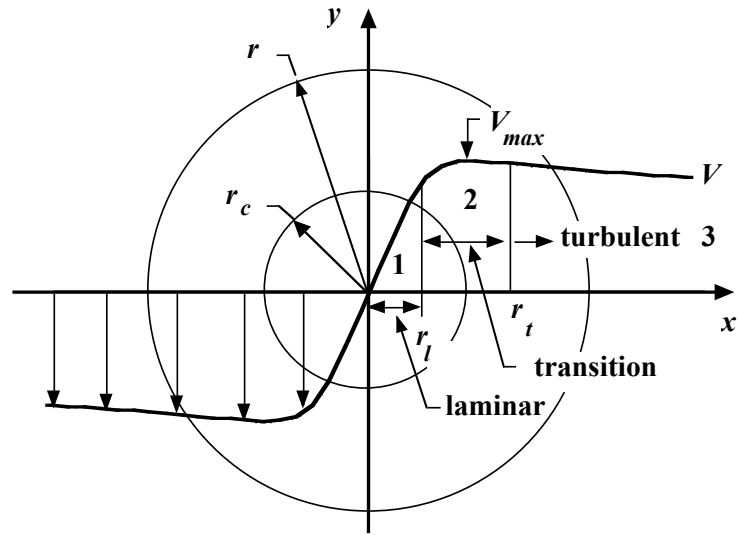

FIGURE 1: Schematic of a turbulent one-cell vortex. The transformation of the flowfield displays the classical characteristics of fast transition to turbulence (see the visualizations in [2] or [9]). In region 1 ($0 < r < r_l$) where the velocity is small and the conditions are laminar-like, the azimuth velocity is seen to increase linearly with the radius. At a radial distance $r_l$, inertia has increased sufficiently to press on the vortex into a transitory region (2) whereby the flow is neither laminar nor turbulent. Full-blown turbulence is reached at $r_t$, occurring always past $V_{max}$, and continues afterwards in region 3.

Nineteen years ago, we reported on a family of laminar vortices [12]; of which the most presently popular member of the set among researchers is the *n* = 2. The basic model was subsequently expanded to account for vortex decay [13], where it was also noted that the system of equations that gave rise to the model was underdetermined and as such there are infinite solutions to the problem. Furthermore, the formulation was yet enlarged to include the effects of density variations [14]. The last extension pointed to the experimental fact that the shape of the tangential velocity remains indifferent to density variations.

In an attempt to capture the overall effects of turbulence, overlooking the finer details, the original *n* = 2 formulation was recently modified [1]:

$$V = \xi \left[ \frac{\alpha+1}{\alpha+\xi^4} \right]^m \qquad (1)$$

The requirement that *V* must be maximum (in fact 1) at $\xi = 1$ yields,

$$m = \frac{\alpha+1}{4}$$

The value of $\alpha$ is determined via the Least-Squares method by minimizing the square error (*E*) of the experimental turbulent vortex profile data:

$$E = \sum_{j=1}^{M} \left\{ V(\xi_j) - \xi_j \left[ \frac{\alpha+1}{\alpha+\xi_j^4} \right]^{\frac{\alpha+1}{4}} \right\}^2 \qquad (2)$$

$V_j$ are the experimental values of the velocity at radial distance $\xi_j$.

Comparable results for $\alpha$ could also be obtained using the experimental far field data of $V_L$ and $\xi_L$ and solving the following equation numerically:

$$4 \ln(\xi_L) + (\alpha+1) \ln\left( \frac{\alpha+1}{\alpha+\xi_L^4} \right) - 4 \ln(V_L) = 0$$

In this formulation the laminar $n = 2$ vortex is recovered when $\alpha = 1$.

Examining Eq. 1 closely one realizes that intense vortices (tangential velocity is dominant) should collapse into a single curve if it is reshaped accordingly:

$$\overline{V} = \frac{1}{1+\varsigma^4} \qquad (3)$$

Where: $\overline{V} = \frac{\alpha}{\alpha+1}\left\{\frac{V}{\xi}\right\}^{\frac{1}{m}}$ and $\varsigma = \frac{\xi}{\alpha^{\frac{1}{4}}}$

The last behavior of the velocity is confirmed in Fig. 3, especially for larger than the core radii (region of present interest) whereby vortices are seen to compress into a single curve irrespective of the vortex size and strength [13], weather they are laminar or not [1], compressible or incompressible [14], steady or time dependent [13]. Therefore, galactic vortices, which are of magnetohydrodynamic nature, belong to the same extended family as their earthly relatives; the mechanically produced and naturally occurring. Another eddy that fits also reasonably well but not included here, is the plasma Burgers-like vortex of Nagaoka et al [15]. Therefore it is reasonable to conclude that the tangential velocity of these vortices remains indifferent to the quality of the underlie medium; i.e. whether it is neutral or ionized fluid.

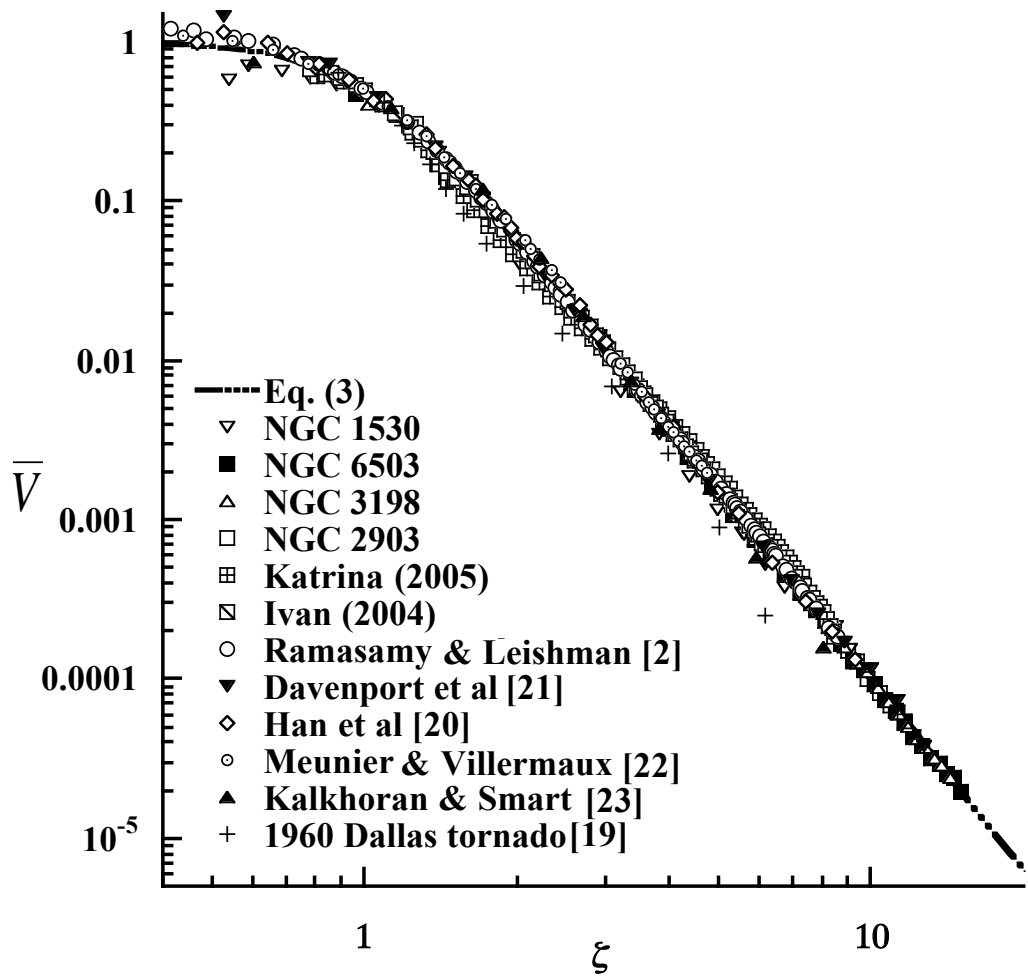

FIGURE 2: Curve-fitted normalized observed velocity distribution with the radius together with Eq. (3) for the extended family of intense vortices. The observations include the tangential velocity curves of: (a) four spiral galaxies; NGC 1530, 6503, 2903 [16] and 3198 [17], (b) the radar data of two tropical cyclones; Katrina (2005) and Ivan (2004) [18], (c) the radar data of the 1960 Dallas tornado [19], (d) the LDA of the laminar Han et al [20] and the Ramasamy and Leishman [2] turbulent helicopter blade vortices, (e) the fixed wing tip vortex of Davenport et al [21], (f) the decaying vortex of Meunier and Villermaux [22], and (g) the compressible vortex of Kalkhoran and Smart [23]. Inside the core, the velocity experiences the maximum change. In addition a vortex being a natural separator causes the number of seed particles to be low thus making the LDA measurements less reliable. These are the reasons as to why the maximum disparity of the observed data from the main curve happens in this region. Also the atmospheric whirls used here were the two-cell vortices of the Sullivan-type. Since the velocity distribution inside the core is not linear but it has a concavity the core data were not included.

Decompression of selected data from Fig. 2 produces Fig. 3. The important to the study elements in Fig. 3 are: (i) all vortices (including the galactic) belong to the same family, (ii) the Ramasamy and Leishman [1] blade vortex is turbulent ($\alpha$ = 0.73), and (iii) the $\alpha$ = 1 condition represents laminar eddies. Therefore, the intensity of turbulence increases with decreasing $\alpha$. By extrapolation, tropical cyclone Bonnie and the three spiral galaxies NGC 1530, 2903, and 3198 are whirls with increasingly turbulence levels making the corresponding vortex velocity to grow successively flat. Note that the most turbulent member in this graph is the spiral galaxy NGC 3198. Actually the expected Reynolds numbers, for these magnetohydrodynamic types of flow are at least 4 orders of magnitude [8] greater than the turbulent helicopter vortices [2].

There is also another common trait in this assemblage of vortices. Based on observations it is a fact that both tropical cyclones and galactic vortices could also experience velocity undulation about the base curve. We hypothesize that this velocity effect is attributed to orbiting waves known to inhabit whirls of this type [25]. Observations made using land-based and airborne radars of several hurricanes revealed that in addition to a circular form of the eyewall in tropical cyclones waveforms due to mesocyclones with wave numbers $N$ = 2, 3, 4, 5, and 6 are also presents [25]-[27]. No heptagonal shape has ever been reported. The last is in agreement with the results of our recent laboratory studies on Kelvin's equilibria [28]. Tornadoes do as well form orbiting secondary revolving eddies known as suction vortices [29]. It is also expected that waves of this nature are present in the interior of spiral galaxies [30]. The observed inner triangular waveforms in the Triangulum galaxy NGC 598, the barred galaxy NGC 613, and the whirlpool galaxy NGC 5194, are revealing [31].

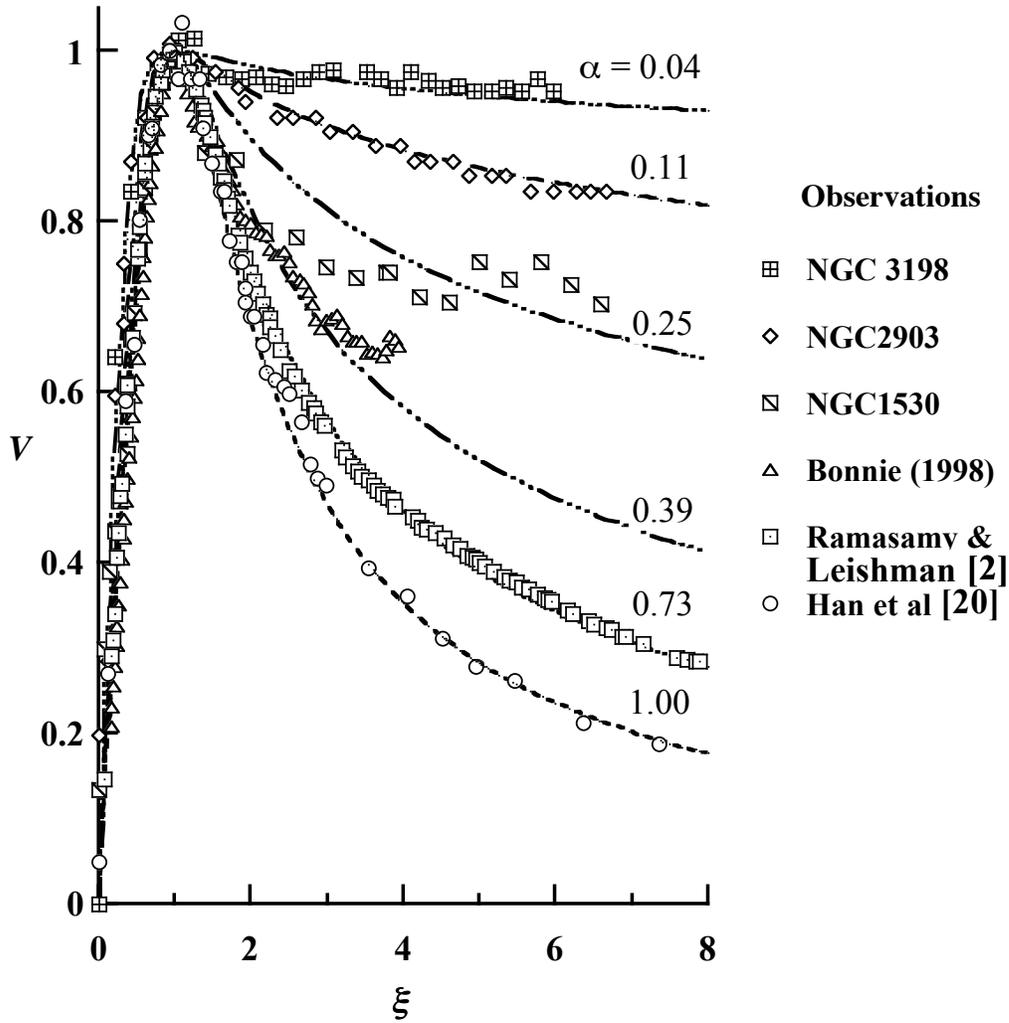

FIGURE 3: Decompressed tangential velocity profiles of selected vortices from Fig. 2 including also the tropical cyclone Bonnie (1998) [24]. The observed LDA data of Ramasamy and Leishman [2] and Han et al [20] tip-vortices are under turbulent and laminar flowfield conditions respectively. The turbulent intensity is thus represented by the curve-fit parameter $\alpha$. When $\alpha = 1$ the original $n = 2$ laminar vortex is recovered while $\alpha \rightarrow 0$ the vortex is under fully turbulent conditions. Therefore, a decreasing value of $\alpha$ represents increase of the turbulent level in the field where the vortex resides.

Superposition of harmonic waves to simulate the last characteristic yields the following approximation for the velocity:

$$V = \frac{\dfrac{b\xi}{\left[\alpha + (b\xi)^4\right]^m} + c\exp(-b\xi)\sin(N\,b\xi)}{\Lambda} \tag{4}$$

Where: $\Lambda = \dfrac{b}{\left[\alpha + (b)^4\right]^m} + c\exp(-b)\sin(N\,b)$, and *a, b* and *c* are scaling constants while *N* is the wave number.

It is evident from Fig. 4 that Eq. (4) can reasonably correlate the flatness of the rotation curve including the accompanied undulations. Based on these results we also speculate that NGC 1530, 2903 and 3198 must have elliptical, triangular and square shaped inner parts respectively.

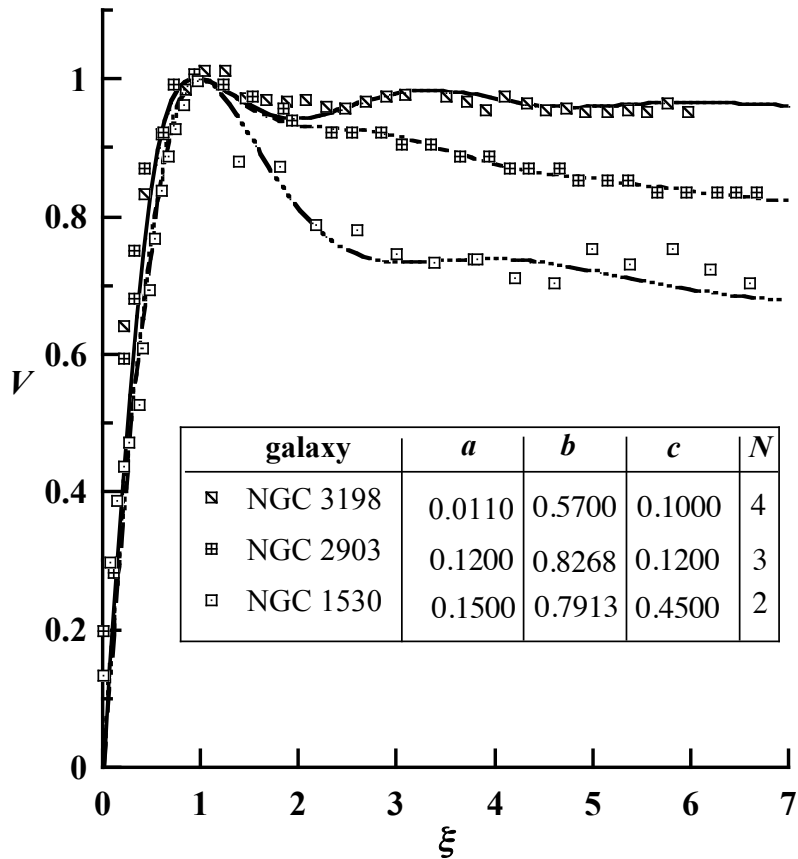

FIGURE 4. Tangential velocity curve-fit using Eq. (4) for three spiral galaxies. The simplistic approach approximates reasonably the uplift and flattening of the individual rotation curves and also the waves present in the medium.

Therefore, based on the previous line of reasoning we conclude that it is indeed possible that intense turbulence in the interstellar gases could explain the apparent flattening in the rotation curves of spiral galaxies.

REFERENCES

[1]   G. H. Vatistas, J. Aircraft, **43**(5) 1577 (2006).
[2]   M . Ramasamy and J. G. Leishman, J. Am. Helicopter Soc. **51** (1) 92 (2006).
[3]   R. B. Larson, Mon. Not. R. Astron. Soc. **186** (2) 479 (1979).
[4]   G. H. Gibsonm, Mech. Rev. **49**(5) 299 (1996).
[5]   R. S. Klessen and P. Hennebelle, Astron. Astrophys., 520, article A17 (2010).
[6]   E. Battaner, J. L. Garrido, M. Membrado and Florido E. Nature **360** 652 (1992).
[7]   Jungyeon Cho and A. Lazarian, Theoret. Comput. Fluid Dynamics **19** 127 (2005).
[8]   A. Lazarian, A. Beresnyak, H. Yan, M. Opher and Y. Liu, Space Sci. Rev. **143** 387 (2009).
[9]   L. D. Landau and E. M. Lifshitz, *Fluid Mechanics* (Pergamon, Oxford, 1987).
[10]  P. B. Martin and J. G. Leishman, Tailing vortex measurements in the wake of a hovering rotor blade with various top shapes. The 58th Annual Forum of the AHS International (2002).
[11]  V. Krishan, Pramana-journal of physics, Vol. **49** (1) 147 (1997).
[12]  G. H. Vatistas, V. Kozel and W. Minh, Exp. Fluids **11** 73 (1991).
[13]  G.H. Vatistas and Y. Aboelkassem, AIAA J.*,* **44**(4) 912 (2006).
[14]  G.H. Vatistas and Y. Aboelkassem, AIAA J.*,* **44** (8) 1912 (2006)
[15]  K. Nagaoka, A. Okamoto, S. Yoshimura, M. Kono, Y. Masayoshi and M. Y. Tanaka, Phys. Rev. Lett. **89**(7), 1075001-1 (2002)
[16]  H. Hoekstra, T. S. van Albada and R. Sancisi, Mon. Not. Astron. Soc. **323**(2) 453 (2001)
[17]  K. G. Begeman, Astron. Astrophys. **223** 47 (1998).
[18]  K. E. Trenberth, C. A. Davis and J. Fasullo. Geophys. Res., **112** D23106 (2007).
[19]  W. H. Hoecker Jr., Monthly Weather Review **88**(5), 167 (1960).
[20]  Y.Q. Han, J.G. Leishman and A.J. Coyone, AIAA J. **35**(3), 477 (1997).
[21]  W. J. Davenport, M. C. Rife, S. I. Liapis and G. J. Follin, J. Fluid Mech*.* **312** 67 (1996).
[22]  P. Meunier and E. Villermaux, J. Fluid Mech*.* **476** 213 (2003).
[23]  I. M. Kalkhoran and M. K. Smart, Prog. Aerosp. Sci. **36**, 63 (2000).
[24]  K. J. Mallen, M. T. Montgomery and B. Wang,. J. Atmos. Sci. **62** 408 (2005).
[25]  W. H. Schubert et al, J. Atmos. Sci.*,* **56**, 1197 (1999).
[26]  B. M. Lewis and H. F. Hawkins, Bull. Am. Meteorol. Soc. **63** 129 (1982).
[27]  T. Muramatsu, J. Meteor. Soc. Japan*,* **64,** 913 (1986).
[28]  G.H. Vatistas, H. Ait-Abderrahmaneand and K. Siddiqui, Phys. Rev. Lett*.* **100**(17) 174503 (2008).
[29]  T. T. Fujita, K. Watanabe, K. Tsuchiya, and M. Shimada, *J. Meteor. Soc. Japan,* **50,** 431 (1972).
[30]  A. M. Fridman, A. G. Morozov, M. V. Nezlin, and E. N. Snezhkin, Physics Letters, **109A** (5), 228 (1985).
[31]  B. C. Elmegreen, D. M. Elmegreen and L. Montenegro, AJ Suppl. Ser. **79** 37 (1992).